\documentclass[11pt,a4paper]{article}
\usepackage[hyperref]{acl2019}
\usepackage{times}
\usepackage{latexsym}
\usepackage{amsmath}
\usepackage{amsfonts}
\usepackage{graphicx}
\usepackage{subcaption}
\usepackage{url}
\usepackage{tabu}
\usepackage{multirow}

\newcommand{\encoder}[2]{\textnormal{#1}^{\textnormal{#2}}}
\newcommand{\equref}[1]{(\ref{#1})}

\aclfinalcopy 

\title{Effective Incorporation of Speaker Information \\in Utterance Encoding in Dialog}

\author{Tianyu Zhao \\
Graduate School of Informatics, \\
Kyoto University \\
{\tt zhao@sap.ist.i.kyoto-u.ac.jp} \\\And
Tatsuya Kawahara \\
Graduate School of Informatics, \\
Kyoto University \\
{\tt kawahara@i.kyoto-u.ac.jp} \\}

\date{}

\begin{document}
\maketitle
\begin{abstract}
    In dialog studies, we often encode a dialog using a hierarchical encoder where each utterance is converted into an utterance vector, and then a sequence of utterance vectors is converted into a dialog vector. Since knowing who produced which utterance is essential to understanding a dialog, conventional methods tried integrating speaker labels into utterance vectors. We found the method problematic in some cases where speaker annotations are inconsistent among different dialogs. A relative speaker modeling method is proposed to address the problem. Experimental evaluations on dialog act recognition and response generation show that the proposed method yields superior and more consistent performances.
\end{abstract}

\section{Introduction}
\label{sec:intro}

In both language understanding and language generation tasks in dialog studies, it is common to encode a dialog as a fixed-length vector, and subsequent processes such as label classification and response generation are conducted on the dialog encoding vector. Since a dialog is often composed of a sequence of utterances, it is convenient to encode each utterance as an utterance vector, and then encode a sequence of utterance vectors as a dialog vector. The hierarchical encoder architecture is effective~\citep{serban2016building}, but it only considers the utterance texts, and does not consider who uttered the words. The lack of speaker information can lead to erroneous predictions in many cases. For example, one usually does not make backchannel to himself/herself. 

Conventional approach to speaker modeling integrates explicit speaker labels into utterance vectors and it works successfully for task-oriented dialogs~\citep{chi2017speaker,chen2017dynamic,kim2019decay}, where only two types of speakers are involved, namely an \textit{Agent} and a \textit{User}. Unfortunately, this approach is not applicable to general dialogs such as the Switchboard corpus~\citep{jurafsky1997switchboard}, because there is a much larger number of speakers and it is impractical to model every speaker separately.\footnote{The number of speaker individuals of the Switchboard Dialog Act corpus~\citep{jurafsky1997switchboard} is not reported, but another version of the corpus (Switchboard-1 Release 2) involves 543 different speakers.} One can convert the speakers into labels such as \textit{A} and \textit{B} for each dialog session and model the two speaker labels only. In this case, two speakers from different sessions may be both annotated as \textit{A} but actually do not refer to the same individual. Such inconsistent annotations can perturb dialog encoding and harm the system performance in subsequent tasks.

To address this problem, we propose to encode the relations between speakers, i.e. whether the speaker of an utterance is identical to the current speaker. By avoiding encoding absolute speaker labels, the proposed method is designed to circumvent inconsistency in speaker annotations. We call the conventional method \textbf{absolute speaker modeling} (Section~\ref{sec:abs}), and the proposed method \textbf{relative speaker modeling} (Section~\ref{sec:rel}). By conducting a systematic comparison between these methods and a baseline without speaker modeling(Section~\ref{sec:exp_set}), we show that proposed method has superior and more consistent performances in two tasks, namely dialog act recognition (Section~\ref{sec:da_recog_task}) and dialog response generation (Section~\ref{sec:generation_task}).

\begin{figure}[!t]
    \centering
    \includegraphics[width=1.0\columnwidth]{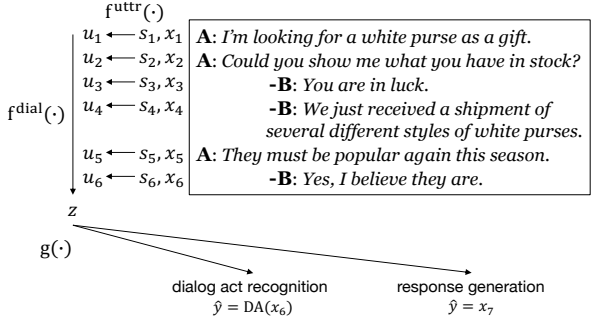}
    \caption{Dialog encoding scheme.}
    \label{fig:dial_encoding}
\end{figure}

\section{Dialog Encoding}

As Figure~\ref{fig:dial_encoding} shows, a dialog $x$ consists of a sequence of utterances, and each utterance $x_{i}$ can be represented by its speaker $s_{i}$ and a sequence of tokens $w_{i,1}, w_{i,2}, \cdots, w_{i,N_{i}}$. For the convenience of further processing, it is common to encode the dialog $x$ as a fixed-length vector $z$ using a dialog encoding function $f(\cdot)$. Then an output function $g(\cdot)$ operates on the dialog vector and predicts an output $\hat{y}$.
\begin{align}
    x &= ((s_{1}, x_{1}), (s_{2}, x_{2}), \cdots, (s_{M}, x_{M}))  \label{equ:def_dial_begin} \\
    x_{i} &= (w_{i,1}, w_{i,2}, \cdots, w_{i,N_{i}}) \label{equ:def_dial_end} \\
    z &= f(x) \\
    \hat{y} &= g(z).
\end{align}

The output function $g(\cdot)$ varies greatly according to the task, but the dialog encoding functions $f(\cdot)$ of most tasks share the same hierarchical encoding architecture. The encoding function can be decomposed into two parts, i.e. an utterance-level encoding function $f^{\textnormal{uttr}}(\cdot)$ that converts an utterance $x_{i}$ to an utterance vector $u_{i}$, and a dialog-level encoding function $f^{\textnormal{dial}}({\cdot})$ that takes a sequence of utterance vectors as input and outputs the dialog vector $z$. We assume that both encoders are implemented by recurrent neural networks (RNNs) since they are widely used in related works. 
\begin{align}
    f(x) &= f^{\textnormal{dial}} ( u_{1}, u_{2}, \cdots, u_{M} ) \\
    u_{i} &= f^{\textnormal{uttr}} (x_{i}) \\
    f^{\textnormal{dial}}(u_{1:M}) &= \encoder{RNN}{dial} (u_{1:M}) \\
    f^{\textnormal{uttr}}(x_{i}) &= \encoder{RNN}{uttr} (x_{i}).
\end{align}

Though the hierarchical architecture provides with a strong baseline encoder, it does not encode speaker information of a dialog into the dialog vector $z$. And the output function $g(\cdot)$ is prone to make incorrect predictions in many cases without knowing the speaker information. 

Therefore, we investigate different approaches to modeling speaker information. We first introduce the \textbf{absolute speaker modeling} method in Section~\ref{sec:abs}, which is adopted by most existing models but suffers from inconsistent speaker annotations in general dialogs. In Section~\ref{sec:rel}, we propose the \textbf{relative speaker modeling} method that yields more a consistent modeling of the speaker role. These methods focus on how to incorporate speaker information into each utterance, and thus only differ in their utterance encoding functions $f^{\textnormal{uttr}} (\cdot)$, while their dialog encoding functions are identical. 

\begin{figure*}[!t]
    \centering
    \includegraphics[width=0.9\textwidth]{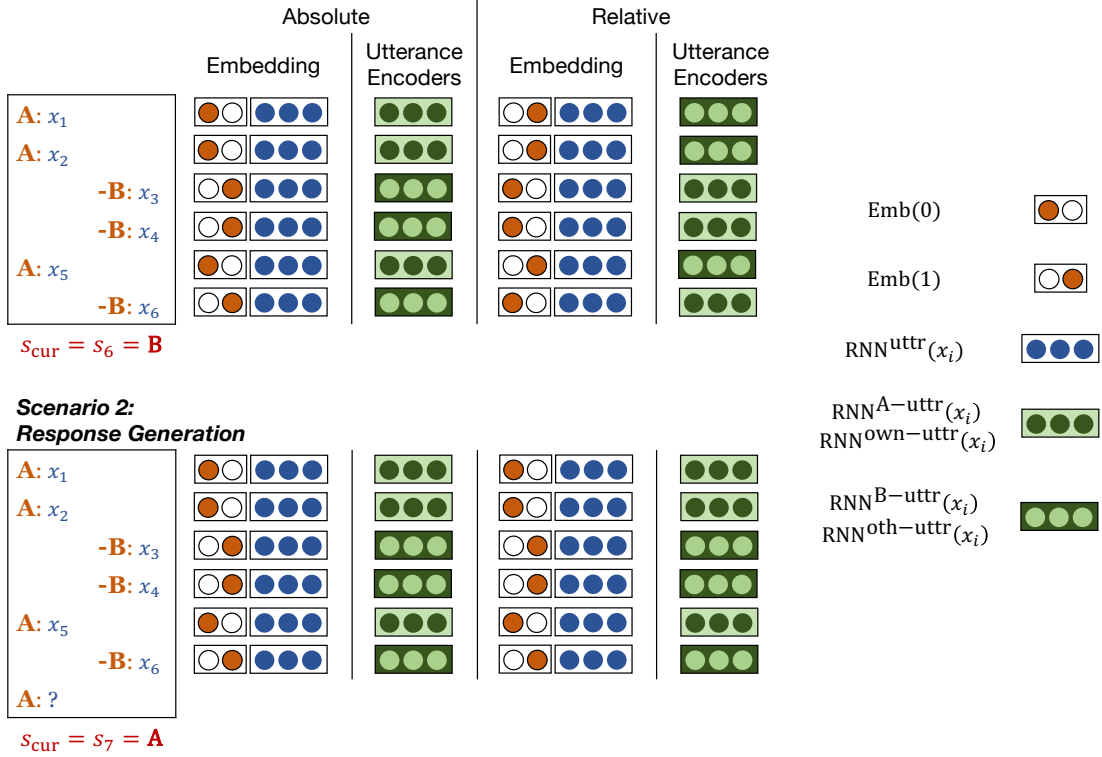}
    \caption{The utterance encodings produced by $f^{\textnormal{uttr}}(\cdot)$ using four speaker modeling methods in two scenarios.}
    \label{fig:scenario}
\end{figure*}

\section{Conventional Method: Absolute Speaker Modeling}
\label{sec:abs}

Speaker information of a dialog $x$ is given by speaker labels $s_{1}, s_{2}, \cdots, s_{M}$. Each speaker label $s_{i}$ is associated with its corresponding utterance $x_{i}$ as shown in Equation~\equref{equ:def_dial_begin}. We assume there are only two speakers \textit{A} and \textit{B} in each dialog,\footnote{The method is also applicable to multi-speaker situations by modification.} and a speaker label $s_{i}$ has value \textit{A} if the utterance $x_{i}$ is produced by speaker \textit{A}, otherwise it has value \textit{B}.

The most straightforward way to incorporate $s_{i}$ is to annotate the utterance vector $u_{i}$ with its speaker, such that $u_{i}$ encodes the information that $x_{i}$ is uttered by \textit{A} or \textit{B}. We call it \textbf{absolute speaker modeling} method, and modify $f^{\textnormal{uttr}}(\cdot)$ to receive both utterance and speaker as arguments:
\begin{align}
    u_{i} &= f^{\textnormal{uttr}} (x_{i}, s_{i}).
\end{align}
We implement the utterance encoding function $f^{\textnormal{uttr}} (\cdot, \cdot)$ using two models, namely \textbf{absolute speaker embedding} and \textbf{absolute speaker utterance encoders}.

\subsection{Absolute Speaker Embedding} 
\label{ssec:abs_emb}
The simplest idea is to use one-hot speaker vectors to encode speaker information, where the speaker vector is $[1,0]$ if the speaker is \textit{A}, otherwise $[0,1]$. We generalize this method to embeddings to obtain speaker vectors. Similar to word embeddings, we define a trainable speaker embedding matrix $\in \mathbb{R}^{2 \times D_{\textnormal{spk}}}$, where $D_{\textnormal{spk}}$ is the speaker embedding dimension. A speaker embedding function $\encoder{Emb}{abs} (s_{i})$ returns the first row of the matrix as the speaker vector if $s_{i}$ is \textit{A}, and the second row if $s_{i}$ is \textit{B}. Then we concatenate the RNN encoding vector $ \encoder{RNN}{uttr} (x_{i} )$ and the speaker embedding vector $\encoder{Emb}{abs} (s_{i})$ to represent the final utterance vector as in following equations.
\begin{align}
    f^{\textnormal{uttr}} (x_{i}, s_{i}) &= \encoder{RNN}{uttr} (x_{i} ) \oplus \encoder{Emb}{abs} (s_{i}) \\
    \encoder{Emb}{abs} (s_{i}) &= 
        \begin{cases} 
            \textnormal{Emb}(0),  & \mbox{if } s_{i} = \mbox{A} \\
            \textnormal{Emb}(1),  & \mbox{if } s_{i} = \mbox{B} \\
        \end{cases},
\end{align}
where $\oplus$ is a concatenation operator.

\subsection{Absolute Utterance Encoders}
\label{ssec:abs_uttr_enc}
In the embedding-based model, an utterance vector is decomposed into two parts, namely the RNN encoding part and the speaker embedding part, and speaker information is all encoded in the speaker embedding. An alternative to absolute speaker embedding is to use different utterance encoders (for example different RNNs), such that speaker information is encoded in the RNN encodings directly by applying different RNNs to utterances from different speakers.

We define two utterance encoders $\encoder{RNN}{A-uttr}$ and $\encoder{RNN}{B-uttr}$, which have the same parameter settings but do not share parameters. The utterance vector $f^{\textnormal{uttr}} (x_{i}, s_{i})$ is given by $\encoder{RNN}{A-uttr}$ if $s_{i}$ is \textit{A}, otherwise it is given by $\encoder{RNN}{B-uttr}$ as following:
\begin{align}
    f^{\textnormal{uttr}} (x_{i}, s_{i}) = 
        \begin{cases} 
            \encoder{RNN}{A-uttr} (x_{i} ),  & \mbox{if } s_{i} = \mbox{A} \\
            \encoder{RNN}{B-uttr} (x_{i} ),  & \mbox{if } s_{i} = \mbox{B} \\
        \end{cases}.
\end{align}

\begin{table*}
    \centering
    \begin{tabu} {c||c|c}
        \textbf{Task} & DA Recognition & Response Generation \\
        \hline
        \hline
        \textbf{Target $y$} & $\textnormal{DA}(x_M)$ & $x_{M+1}$ \\
        \hline
        \textbf{Output function $g(\cdot)$} & Fully-connected layer & RNN decoder \\
        \hline
        \textbf{Corpus} & Switchboard Dialog Act Corpus & DailyDialog Corpus \\
        \hline
        \textbf{\# of sessions (train/dev/test)} & 1,003/112/19 & 9,031/1,127/1,134 \\
        \hline
        \textbf{\# of utterances} & 178k/18k/4k & 104k/17k/14k \\
        \hline
        \textbf{\# of tokens} & 1,730k/180k/40k & 935k/146k/126k \\
        \hline
        \textbf{\# of DA labels} & 42 & \textit{Not Used}
        
    \end{tabu}
    \caption{Statistics of two corpora and three tasks for evaluation.}
    \label{tab:stat}
\end{table*}

\section{Proposed Method: Relative Speaker Modeling}
\label{sec:rel}

Many works have shown that the absolute speaker modeling method is able to capture speaker-specific patterns to some extent, and it improves the performance of baseline models that do not incorporate speaker information~\citep{chi2017speaker,chen2017dynamic,kim2019decay}. These works focused on task-oriented dialogs, where two types of speakers are involved, namely an \textit{Agent} and a \textit{User}. The roles of the two speakers are consistent and clearly defined, where the \textit{User} gives requests to complete a certain task and the \textit{Agent} provides appropriate actions. And the speaker embeddings or the utterance encoders are able to learn role-specific information easily. In general dialogs, however, there are a much larger number of speaker individuals. If we simply annotate speakers as labels (\textit{A}, \textit{B}, etc.) for each session, two \textit{A}s from different sessions may not refer to the same real individual. Such inconsistency in speaker annotations can degrade the performances of the absolute speaker models. 

To avoid using absolute speaker annotations, we propose to encode the information ``the relation between speaker $s_{i}$ and the current speaker $s_{\textnormal{cur}}$'', instead of ``speaker $s_{i}$ is A or B'' as in the absolute speaker models.

\textbf{Definition of Current Speaker.} The current speaker $s_{i}$ depends on a scenario, for example in Scenario A in Figure~\ref{fig:scenario}, the task is to recognize the dialog act of the last utterance $x_{6}$, thus the current speaker is $s_{6}$. And in Scenario B, where the task is to predict the next utterance as a response, the current speaker becomes $s_{7}$. 

\textbf{Modeling of Speaker Relation.} To encode the speaker relation between $s_{i}$ and $s_{\textnormal{cur}}$, we annotate the utterance vector $u_{i}$ with the information of whether $s_{i}$ equals to $s_{\textnormal{cur}}$ or not. One can interpret the information as whether the utterance $x_{i}$ is my own utterance or others' utterance, and the model becomes \textit{self}-aware by defining \textit{self} as $s_{\textnormal{cur}}$.

Since only the relation between speakers is encoded in the utterance vector $u_{i}$, we call it a \textbf{relative speaker modeling} method. We extend the utterance encoding function $f^{\textnormal{uttr}}$ to include the current speaker $s_{\textnormal{cur}}$ as an argument:
\begin{align}
    u_{i} &= f^{\textnormal{uttr}} (x_{i}, s_{i}, s_{\textnormal{cur}}).
\end{align}
Models using absolute speaker modeling can be easily converted into relative speaker modeling ones. We illustrate two models as counterparts to the absolute speaker embedding model and absolute speaker utterance encoders model mentioned in the last section.

\subsection{Relative Speaker Embedding}
The absolute speaker embedding model in Section~\ref{ssec:abs_emb} is converted into a relative speaker model by redefining its speaker embedding function. The new embedding function $\encoder{Emb}{rel} (s_{i}, s_{\textnormal{cur}})$ now returns one embedding when $s_{i}$ is identical to $s_{\textnormal{cur}}$, otherwise another embedding.
\begin{align}
    f^{\textnormal{uttr}} (x_{i}, s_{i}, s_{\textnormal{cur}}) &= \encoder{RNN}{uttr} (x_{i} ) \oplus \encoder{Emb}{rel} (s_{i}, s_{\textnormal{cur}}) \\
    \encoder{Emb}{rel} (s_{i}, s_{\textnormal{cur}}) &= 
        \begin{cases} 
            \textnormal{Emb}(0),  & \mbox{if } s_{i} = s_{\textnormal{cur}} \\
            \textnormal{Emb}(1),  & \mbox{otherwise} \\
        \end{cases}.
\end{align}

\subsection{Relative Utterance Encoders}
Similar to the absolute utterance encoders model in Section~\ref{ssec:abs_uttr_enc}, a relative utterance encoders model also has two utterance encoders $\encoder{RNN}{own-uttr}$ and $\encoder{RNN}{oth-uttr}$. In contrast to its counterpart, however, it uses $\encoder{RNN}{own-uttr}$ to encode an utterance $x_{i}$ if its speaker $s_{i}$ is the same as $s_{\textnormal{cur}}$, otherwise it uses $\encoder{RNN}{oth-uttr}$.
\begin{align}
    \begin{split}
    f^{\textnormal{uttr}} (x_{i}, &s_{i}, s_{\textnormal{cur}}) = \\
        &\begin{cases} 
            \encoder{RNN}{own-uttr} (x_{i} ),  & \mbox{if } s_{i} = s_{\textnormal{cur}} \\
            \encoder{RNN}{oth-uttr} (x_{i} ),  & \mbox{otherwise} \\
        \end{cases}.
    \end{split}
\end{align}

A comparison between outputs of the two absolute speaker models and the two relative speaker models is illustrated in Figure~\ref{fig:scenario}. Using the same dialog as an example, we define two scenarios where the tasks are different, such that the current speakers $s_{\textnormal{cur}}$ are different. Since the absolute speaker labels remain the same in two tasks, the utterance vectors yielded by the absolute speaker models do not change. On the other hand, the outputs of the relative speaker models vary in the two scenarios because the change of $s_{\textnormal{cur}}$ leads to the change of speaker relations.

\section{Experiment Settings}
\label{sec:exp_set}

\begin{table*}  
    \centering
    \begin{tabu} {l||c|c|c}
        \multicolumn{1}{c|}{\textbf{Model}} & \textbf{Macro F1 $\uparrow$} & \textbf{Weighted F1 $\uparrow$} & \textbf{Accuracy $\uparrow$} \\
        \hline
        \hline
        \textit{Baseline} & & & \\
        \quad - without speaker modeling & 54.27 & 78.20 & 79.00 \\
        \hline
        \textit{Absolute Speaker Modeling} & & & \\
        \quad - embedding & 52.40 & 77.53 & 78.70 \\
        \quad - utterance encoders & 55.33 & 77.97 & 78.90 \\
        \hline
        \textit{Relative Speaker Modeling} & & & \\
        \quad - embedding & 54.90 & 78.70 & 79.63 \\
        \quad - utterance encoders & \textbf{57.90} & \textbf{79.50} & \textbf{80.17}$^\star$ \\
        \hline
        \hline
        \textit{Past Works} & & & \\
        \quad \citet{wan2018improved} & & & 81.5 \\
        \quad \citet{raheja2019dialogue} & & & 82.9$^\dagger$ \\
        \hline
        \textit{Annotator Inter-agreement} & & & 84.0 
    \end{tabu}
    \caption{Results of DA recognition. $^\star$ The result is significantly better than the baseline with $p$-value $< 0.001$.} 
    \label{tab:res_recog}
\end{table*}

\subsection{Tasks and Corpora}
For assessment of the speaker modeling methods, we conduct experiments on two tasks, namely dialog act (DA) recognition and response generation, using two corpora, namely the Switchboard Dialog Act (SwDA) corpus~\citep{jurafsky1997switchboard} and DailyDialog (DD) corpus~\citep{li2017dailydialog}.

\textbf{Dialog Act Recognition.} Given a dialog $x$ of $M$ utterances, a DA recognition model classifies the dialog act of the last utterance in the dialog, i.e. $\textnormal{DA}(x_M)$. A hierarchical encoding function $f(\cdot)$ encodes $x$ into $z$, in which the utterance encoding function $f^{\textnormal{uttr}}(\cdot)$ varies according to speaker modeling method. Then we add a simple fully-connected layer as the output function $g(\cdot)$ to predict the probabilities of DA labels. The entire model is trained to maximize likelihood $p(\textnormal{DA}(x_M)|x)$.


\textbf{Response Generation.} A response generation model predicts the upcoming utterance $x_{M+1}$. Given a dialog $x$ as context, a decoder RNN initializes its hidden states with the dialog vector $z$. Then it decodes a hypothesis utterance $\hat{x}_{M+1}$ word by word. We optimize the model to maximize $p(x_{M+1}|x)$.

A summary of the tasks and used corpora is given in Table~\ref{tab:stat}. 

\subsection{Implementation}
We experiment with (1) a baseline model without speaker modeling, (2) the absolute speaker embedding model, (3) the absolute utterance encoders model, (4) the relative speaker embedding model, and (5) the relative utterance encoders model. The hyperparameters are given as following. We keep the most frequent 10,000 words as our vocabulary. Word embeddings are 200-dimensional and initialized with the pre-trained \textit{Glove}~\citep{pennington2014glove} embeddings. All the utterance encoder RNNs are implemented by one-layer bidirectional gated recurrent unit networks~\citep{cho2014gru} and have 300 hidden units. Dialog encoder RNNs are unidirectional instead. Speaker embeddings are randomly initialized and the embedding size $D_{\textnormal{spk}}$ is 30. The decoder RNN in response generation task is a one-layer and unidirectional GRU network with 300 hidden units. All the models are trained for 20 epochs with a batch size 30 by the Adam method~\citep{kingma2014adam}. The learning rate is initialized as $1 \times 10^{-3}$ and halves when validation loss does not decrease. Early stopping is applied when the learning rate is smaller than $1 \times 10^{-7}$.

\begin{figure*}[!t]
    \centering

    \begin{subfigure}{0.32\textwidth}
    \centering
    \includegraphics[width=\linewidth]{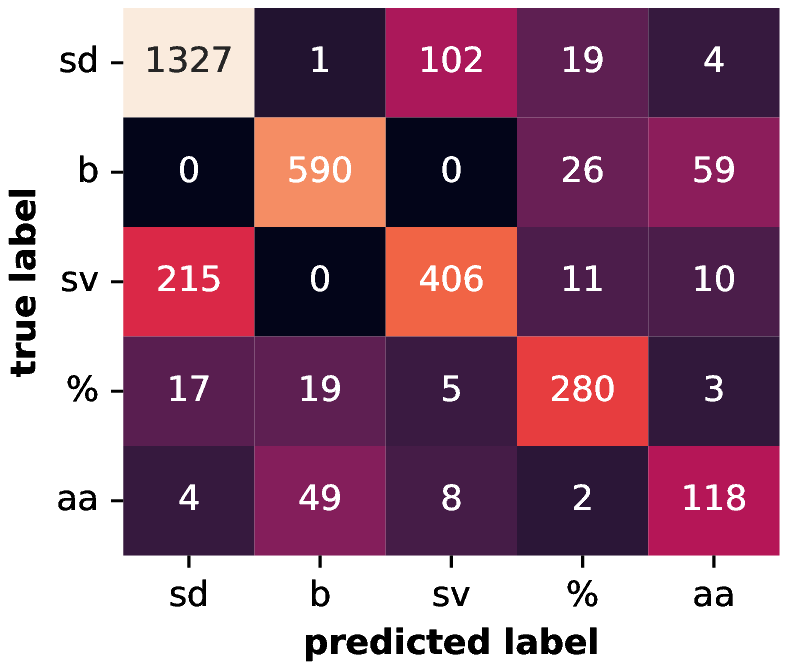}
    \caption{Without speaker modeling.}
    \label{fig:conf_mat_recog_wo_speaker}
    \end{subfigure}
    \begin{subfigure}{0.32\textwidth}
    \centering
    \includegraphics[width=\linewidth]{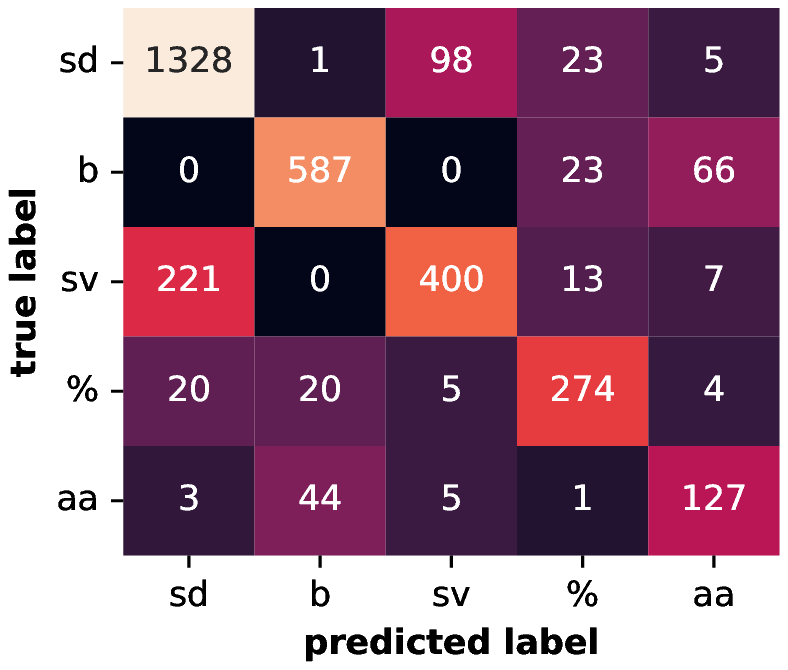}
    \caption{Absolute utterance encoder.}
    \label{fig:conf_mat_recog_abs_enc}
    \end{subfigure}
    \begin{subfigure}{0.32\textwidth}
    \centering
    \includegraphics[width=\linewidth]{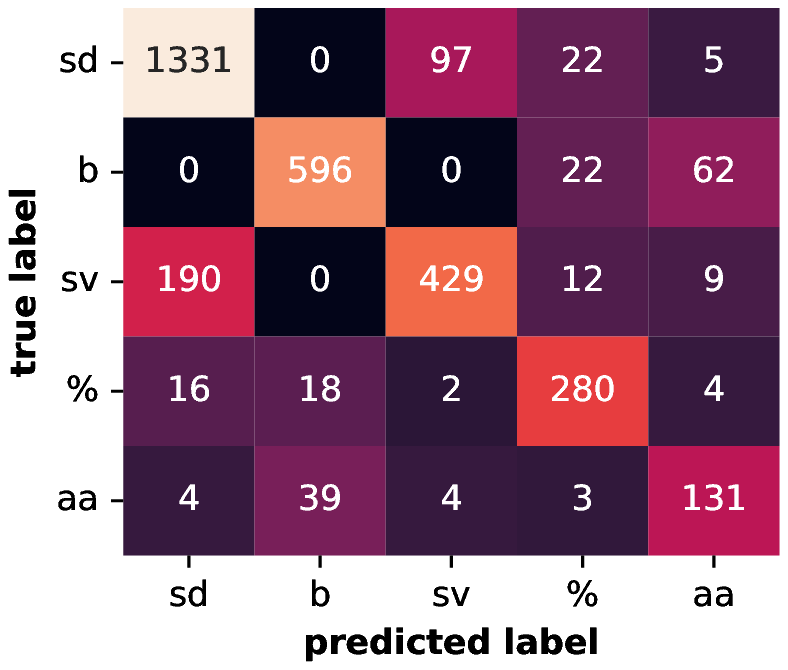}
    \caption{Relative utterance encoder.}
    \label{fig:conf_mat_recog_rel_enc}
    \end{subfigure}
    
    \caption{Confusion matrices of top-5 frequent DA labels of three models in DA recognition task.}
    \label{fig:conf_mat_recog}
\end{figure*}

\begin{table*}[!t]
    \centering
    \begin{tabu} {c|c|c|c|c|c}
        \textbf{Dialog Act} & \multirow{2}{*}{\textbf{Speaker}} & \multicolumn{2}{c|}{\textbf{Absolute Model}} & \multicolumn{2}{c}{\textbf{Relative Model}} \\
        \cline{3-6}
        \textbf{(\# examples)} & & \textbf{Accuracy $\uparrow$} & \textbf{$\Delta \downarrow$} & \textbf{Accuracy $\uparrow$} & \textbf{$\Delta \downarrow$} \\
        \hline
        \hline
        \multirow{2}{*}{sd (1481)} & A & 90.36 & \multirow{2}{*}{1.35} & 89.39 & \multirow{2}{*}{\textbf{0.94}} \\
        \cline{2-3} \cline{5-5} 
        & B & 89.01 & & 90.33 & \\
        \hline
        \multirow{2}{*}{b (696)} & A & 84.62 & \multirow{2}{*}{\textbf{0.54}} & 86.10 & \multirow{2}{*}{0.90} \\
        \cline{2-3} \cline{5-5} 
        & B & 84.08 & & 85.29 & \\
        \hline
        \multirow{2}{*}{sv (659)} & A & 58.77 & \multirow{2}{*}{4.01} & 64.91 & \multirow{2}{*}{\textbf{0.39}} \\
        \cline{2-3} \cline{5-5} 
        & B & 62.78 & & 65.30 & \\
        \hline
        \multirow{2}{*}{\% (327)} & A & 85.80 & \multirow{2}{*}{3.98} & 85.80 & \multirow{2}{*}{\textbf{0.34}} \\
        \cline{2-3} \cline{5-5} 
        & B & 81.82 & & 85.46 & \\
        \hline
        \multirow{2}{*}{aa (187)} & A & 65.93 & \multirow{2}{*}{3.86} & 68.13 & \multirow{2}{*}{\textbf{3.75}} \\
        \cline{2-3} \cline{5-5} 
        & B & 69.79 & & 71.88 &
    \end{tabu}
    \caption{Comparison of speaker-specific accuracies of top-5 frequent DA labels in DA recognition task.}
    \label{tab:dist_diff_recog}
\end{table*}

\section{Task 1: Dialog Act Recognition}
\label{sec:da_recog_task}

\subsection{Evaluation Results}
For evaluation of DA recognition models, we calculate their macro (unweighted) F1 scores, weighted F1 scores, and accuracies, which are common measurements in multi-class classification tasks. Results of the five models are reported in Table~\ref{tab:res_recog}. Surprisingly, the two absolute speaker models perform even slightly worse than the baseline without speaker modeling. The relative speaker models outperform the baseline and their absolute counterparts. The relative utterance encoders model gives significant improvements over the baseline in all measurements (+3.63\% macro F1, +1.3\% weighted F1, and +1.17\% accuracy). We use McNemar's test, an non-parametric test for paired binary data, to assess the significance of accuracy improvement. It shows that the improvement is significant with $p$-value $4.86 \times 10^{-4}$. The improvements in F1 measures are comparable to or larger than the improvement in accuracy.

Table~\ref{tab:res_recog} shows state-of-the-art results from previous works. \citet{wan2018improved} proposed to use a question-answering network with dynamic memory and reached 81.5\% accuracy on the SwDA corpus. \citet{raheja2019dialogue} reached 82.9\% accuracy, the best score reported so far. They applied hierarchical RNN encoders with self attention and a CRF layer to decode the optimal DA sequence of an entire dialog, so the model has access to information from future utterances. We do not aim to surpass these results because they exploit much more complex model architectures. However, neither of the works used any speaker features. We can hopefully further improve the models' performances by applying the relative speaker modeling method.

\subsection{Analysis}

To understand how relative speaker modeling improves DA recognition accuracy, we compare the confusion matrices of the baseline model, the absolute utterance encoders model, and the relative utterance encoders model. As shown in Figure~\ref{fig:conf_mat_recog}, one major difference is that the relative model (Figure~\ref{fig:conf_mat_recog_rel_enc}) performs better in distinguishing \textsc{Statement-opinion}(sv) from \textsc{Statement-non-opinion}(sd). 215 and 221 \textsc{Statement-opinion} utterances are misclassified as \textsc{Statement-non-opinion} by the baseline model and the absolute model, respectively. The relative model reduces the number to 190. 


\begin{table*}[!t]
    \centering
    \begin{tabu} {l||c|c|c|c|c}
        \multicolumn{1}{c|}{\textbf{Model}} & \textbf{BLEU-1 $\uparrow$} & \textbf{BLEU-2 $\uparrow$} & \textbf{SIF emb. $\uparrow$} & \textbf{Distinct-1 $\uparrow$} & \textbf{Distinct-2 $\uparrow$} \\
        \hline
        \hline
        \textit{Baseline} & & & & & \\
        \quad - without speaker modeling & 14.94 & 2.10 & 11.21 & 1610 & 6999 \\
        \hline
        \textit{Absolute Speaker Modeling} & & & & & \\
        \quad - embedding & 14.86 & 2.06 & 9.48 & 1425 & 5917 \\
        \quad - utterance encoders & 14.48 & 1.92 & 10.68 & 1579 & 7047 \\
        \hline
        \textit{Relative Speaker Modeling} & & & & & \\
        \quad - embedding & 15.66$^\star$ & 2.37$^\star$ & 16.87$^\star$ & 1468 & 6409 \\
        \quad - utterance encoders & \textbf{15.72}$^\star$ & \textbf{2.52}$^\star$ & \textbf{18.18}$^\star$ & \textbf{1752}$^\star$ & \textbf{8266}$^\star$
    \end{tabu}
    \caption{Results of response generation. $^\star$ The result is significantly better than the baseline with $p$-value $< 0.001$.}
    \label{tab:res_generation}
\end{table*}

We reason that the absolute speaker models performed worse because they learnt speaker biases (e.g. different label distributions) of the training data set. Since speaker annotation is inconsistent in the corpus, the speaker biases are different in the test data set. If so, there should be a large difference between the absolute model's accuracies on \textit{A}'s utterances and \textit{B}'s utterances. On the other hand, the relative model should have more consistent performance. Therefore, we compare the DA- and speaker-specific accuracies of the absolute utterance encoders model and the relative utterance encoders model in Table~\ref{tab:dist_diff_recog}. Let $\Delta$ denotes the difference between speaker-specific accuracies. The relative model generally has lower $\Delta$ numbers. Specifically, the relative model performs much more consistently in recognizing \textsc{Statement-opinion}(sv) and \textsc{Abandoned or Turn-Exit}(\%).

\section{Task 2: Response Generation}
\label{sec:generation_task}

\subsection{Evaluation Results}
In the task of dialog response generation, we evaluate the relevance and diversity of responses generated by the models on DD corpus. To assess relevance, we use two $n$-gram overlapping scores, i.e. BLEU-$1$ and BLEU-$2$~\citep{papineni2002bleu}, and an embedding similarity score called smooth inverse frequency (SIF) embedding similarity. For diversity evaluation, we calculate the number of distinct $n$-gram types, denoted as Distinct-$n$~\citep{li2016diversity}.

\textbf{SIF embedding similarity.} Commonly used embedding similarity scores such as embedding average have been shown to correlate weakly with human judgement. One of the reasons is that a large number of tokens (for example function words) contribute to the resulting sentence embeddings, but have little effect on semantic relevance. To mitigate such noise, we adapt the SIF embedding proposed by \citet{arora2017sif} to compute sentence embeddings. An SIF sentence embedding is the weighted average of the word embeddings in a sentence, with its projection on a first principal component subtracted. The weight of a word $w$ is $\frac{a}{a+p(w)}$, where $a$ is a parameter and $p(w)$ is the word frequency estimated from the training corpus. The first principal component is computed from all generated responses. Given SIF embedding of a reference utterance and a hypothesis utterance, we calculate the cosine similarity between the two embedding vectors.

In Table~\ref{tab:res_generation}, we compare the five models on five measurements, namely BLEU-$1$, BLEU-$2$, SIF embedding similarity, Distinct-$1$, and Distinct-$2$. The results show the same trend as in DA recognition. Two absolute models perform worse than the baseline, and the relative models have better performance in all five measurements. The relative utterance encoders model reaches the highest scores in both relevance and diversity metrics. Improvement in SIF embedding similarity is especially large compared to the baseline (6.97\% absolute increase and 62.18\% relative increase). For significance test, we use the Wilcoxon signed-rank test, a non-parametric test for paired data. All the scores of the relative utterance encoders model are significantly better than the baseline.

\begin{table*}[!t]
    \centering
    \begin{tabu} {c|c|c|c|c|c}
        \multirow{2}{*}{\textbf{Metric}} & \multirow{2}{*}{\textbf{Speaker}} & \multicolumn{2}{c|}{\textbf{Absolute Model}} & \multicolumn{2}{c}{\textbf{Relative Model}} \\
        \cline{3-6}
         & & \textbf{Score $\uparrow$} & \textbf{$\Delta \downarrow$} & \textbf{Score $\uparrow$} & \textbf{$\Delta \downarrow$} \\
        \hline
        \hline
        \multirow{2}{*}{BLEU-$1$} & A & 15.48 & \multirow{2}{*}{1.69} & 16.31 & \multirow{2}{*}{\textbf{1.28}} \\
        \cline{2-3} \cline{5-5} 
        & B & 13.79 & & 15.04 & \\
        \hline
        \multirow{2}{*}{BLEU-$2$} & A & 3.04 & \multirow{2}{*}{\textbf{0.51}} & 3.65 & \multirow{2}{*}{0.58} \\
        \cline{2-3} \cline{5-5} 
        & B & 2.53 & & 3.07 & \\
        \hline
        \multirow{2}{*}{SIF emb.} & A & 13.10 & \multirow{2}{*}{1.99} & 18.60 & \multirow{2}{*}{\textbf{0.11}} \\
        \cline{2-3} \cline{5-5} 
        & B & 11.11 & & 18.71 & \\
        \hline
        \multirow{2}{*}{Distinct-$1$} & A & 1362 & \multirow{2}{*}{150} & 1447 & \multirow{2}{*}{176} \\
        \cline{2-3} \cline{5-5} 
        & B & 1212 & & 1271 & \\
        \hline
        \multirow{2}{*}{Distinct-$2$} & A & 5418 & \multirow{2}{*}{732} & 6195 & \multirow{2}{*}{1002} \\
        \cline{2-3} \cline{5-5} 
        & B & 4686 & & 5193 &
    \end{tabu}
    \caption{Comparison of speaker-specific metrics in response generation task.}
    \label{tab:dist_diff_generation}
\end{table*}

\subsection{Analysis}
Similar to the analysis in DA recognition task, we compare the speaker-specific scores between the absolute and the relative utterance encoders models in Table~\ref{tab:dist_diff_generation}. The relative model performs more consistently according to its speaker-specific BLEU-$1$ scores and SIF embedding similarities, while the two models' consistencies in BLEU-$2$ scores are comparable. As with diversity scores, we should first take into account the number of $A$'s utterances and that of $B$'s. The ratio of \textit{A}'s to \textit{B}'s utterances is 1.29 (7201/5575) in the test data set. The relative model's Distinct-$1$ ratio is 1.14 (1447/1271) and its Distinct-$2$ ratio is 1.19 (6195/5193), both slightly closer to 1.29 than the absolute model's 1.12 (1362/1212) and 1.16 (5418/4686).



\section{Related Works}
\label{sec:related}

\textbf{Absolute Speaker Modeling.} Speaker-specific utterance encoders are common in task-oriented natural language understanding~\citep{chi2017speaker,chen2017dynamic,kim2019decay}. They used separate convolutional neural networks (CNNs) to implement the utterance encoding function $f^{\textnormal{uttr}} (x_{i}, s_{i})$. Their absolute speaker modeling methods achieved considerable improvements compared to baselines without speaker modeling because task-oriented dialogs have a consistent annotation of speakers, namely \textit{Agent} and \textit{User}. In dialog response generation task,~\citet{shen2017conditional} proposed another absolute speaker modeling method. They modified the hierarchical recurrent encoder decoder (HRED) proposed by~\citet{serban2016building}, and used separate dialog-level RNN encoders for different speakers, which can be considered as a variation of the absolute utterance encoders model in this paper.

\textbf{Relative Speaker Modeling.} The relative utterance encoders model in Section~\ref{sec:rel} has a similar architecture compared to the speaker interaction RNNs proposed by~\citet{zhang2018addressee}. They modeled speaker status for all speakers in a multi-speaker conversation. Each speaker status is a fixed-length vector and gets updated by a shared dialog-level RNN and three types of utterance encoder RNNs, namely the \textit{sender} RNN, the \textit{addressee} RNN, and the \textit{observer} RNN. The first two types of RNNs resemble our $\encoder{RNN}{own-uttr}$ and $\encoder{RNN}{oth-uttr}$, while the \textit{observer} RNN is designed for multi-speaker scenario. Though the resulting models are similar, we point out three major differences between their work and ours. (1) We derive the relative utterance encoders model from its counterpart that uses absolute speaker modeling, and we have made a systematic comparison between absolute speaker models and relative speaker models. (2) The speaker interaction RNN introduced a role-specific RNN gate that incorporates outputs from RNNs of other roles, and thus makes the model complex. Our proposal, however, requires minimal modification to the baseline and the absolute speaker model, while yielding significant improvements. (3) They focused on next speaker and utterance selection, and our evaluation experiments show that the relative speaker modeling method is beneficial to both language understanding and language generation. 

\textbf{Speaker Identity Modeling.} Another line of speaker modeling in dialog is speaker identity modeling. Especially in dialog response generation, speaker individuals' information (e.g. speaker profile and speaking style) is essential for generating consistent and truth-grounding responses. Prior works have proposed to model speaker identity by exploiting given profiles~\citep{zhang2018personalizing}, clustered speaker embeddings~\citep{li2016persona}, and dynamically computed speaker feature vectors~\citep{zhang2019consistent}. We also look forward to integrating speaker identity information into the proposed architecture in the future.

\section{Conclusion}
We investigated speaker modeling methods in dialog encoding, and found that conventional absolute speaker modeling method sometimes suffers from the inconsistent speaker annotations of a corpus. To avoid the inconsistency, we proposed a relative speaker modeling method, which encodes relations between speakers instead of speaker labels themselves. We implemented both methods with embeddings and separate utterance encoders, and conducted a systematic comparison between the four models and a baseline without speaker modeling. Experimental results on two tasks and two corpora demonstrated that the relative speaker modeling method shows consistent improvements over others.

\bibliography{../collection}
\bibliographystyle{acl_natbib}

\appendix

\end{document}